# Infrared complex refractive index of astrophysical ices exposed to cosmic rays simulated in the laboratory

W. R. M. Rocha,[1]✕ S. Pilling,[1] A. L. F. de Barros,[2] D. P. P. Andrade,[3] H. Rothard[4] and P. Boduch[4]

[1]*Instituto de Pesquisa e Desenvolvimento (IP&D), Universidade do Vale do Paraíba, Av. Shishima Hifumi 2911, CEP 12244-000, São José dos Campos, SP, Brazil*
[2]*Departamento de Física, Centro Federal de Educação Tecnológica Celso Suckow da Fonseca, Av. Maracanã 229, 20271-110 Rio de Janeiro, RJ, Brazil*
[3]*Observatório do Valongo, Universidade Federal do Rio de Janeiro - UFRJ, Rio de Janeiro, Brazil*
[4]*Centre de Recherche sur les Ions, les Matériaux et la Photonique, Normandie Univ, ENSICAEN, UNICAEN, CEA, CNRS, CIMAP, 14000 Caen, France*

**ABSTRACT**

In dense and cold regions of the interstellar medium (ISM), molecules may be adsorbed onto dust grains to form the ice mantles. Once formed, they can be processed by ionizing radiation coming from stellar or interstellar medium leading to formation of several new molecules in the ice. Among the different kind of ionizing radiation, cosmic rays play an important role in the solid-phase chemistry because of the large amount of energy deposited in the ices. The physicochemical changes induced by the energetic processing of astrophysical ices are recorded in a intrinsic parameter of the matter called complex refractive index (CRI). In this paper, we present for the first time a catalogue containing 39 complex refractive indices (n, k) in the infrared from 5000 and 600 cm$^{-1}$ (2.0 - 16.6 μm) for 13 different water-containing ices processed in laboratory by cosmic ray analogs. The calculation was done by using the NKABS (acronym of determination of N and K from ABSorbance data) code (Rocha & Pilling 2014), which employs the Lambert-Beer and Kramers-Kronig equations to calculate the values of n and k. The results are also available at the website: http://www1.univap.br/gaa/nkabs-database/data.htm. As test case, a $H_2O:NH_3:CO_2:CH_4$ ice was employed in a radiative transfer simulation of a protoplanetary disk to show that these data are indispensable to reproduce the spectrum of YSOs containing ices.

# 1 INTRODUCTION

In the coldest regions of the ISM (T < 20 K), most of the molecules in the gas-phase can be adsorbed onto dust grains, and the column density can be high enough to increase the visual extinction to values greater than 3 mag (Whittet et al.1974;Öberg et al.2009). In a typical scenario, according toÖberg et al.(2009), the hydrogenation mechanism of the adsorbed atoms leads to the formation of simple ices such as $H_2O$, CO, $CO_2$, $CH_4$, and $NH_3$. Such a scenario is common in dense molecular clouds, and also possible in the initial stages of protostars (also known as young stellar objects – YSOs), called class 0 or class I according to Lada's classification (Lada et al. 1987). However, recent laboratory experiments have shown that complex molecules, like hydroxylamine ($NH_2OH$), can also be formed on dust grains at low temperature by hydrogenation of NO ice without external energetic input (Congiu et al.2012a), or via ammonia oxidation (He et al. 2015). Although these two mechanisms are able to explain the formation of $NH_2OH$ at low temperature,He et al.(2015) point out that its formation can be explained, in a better approximation, by ammonia oxidation once the abundance of $NH_3$ is larger than that of NO in the interstellar medium. In this case, $NH_3OH$ can be formed around 14 K, when O atoms move around a potential well where $NH_3$ is adorbed, or even around 70 K,when O atoms can interact with $NH_3$ by Eley-Rideal or hot-atom mechanisms.

Several astronomical observations report the presence of ices indicated by strong absorptions in the near- and mid-infrared spectra of low- and high-mass YSOs (Boogert et al.2000,2002a,2008; Pontoppidan et al.2008;Öberg et al.2008;Rocha et al.2015), in dense molecular clouds (Knez et al.2008;Hollenbach et al.2009), and in the vicinity of AGNs - Active Galactic Nuclei (Spoon et al.2002). Since the formation of ices is common in the ISM, it is important to understand the role of the ionizing radiation field in the solid-phase chemistry. Several studies were performed to understand the formation of complex molecules like $NH_4^+OCN^-$ and $NH_3^+CH_2COO^-$ (Pilling et al.2010a;Schutte et al.2003), $HCOO^-$ and $H_2CO_3$ (Pilling et al.2010b), $CH_3OH$ (Öberg et al. 2009;Andrade et al.2014), $C_2N_2$, $HC_3N$ and $CH_3CN$ (Moore et al.2010;Hudson et al.2014), $HOCH_2CHO$ and $HOCH_2CH_2OH$ (Butscher et al.2015). In the same way, other investigations addressed the absorption features in the protostellar Spectral Energy Distribution (SED) by using the mix-and-match scheme to fit the ice absorption profiles observed in the protostellar spectra, although such a procedure is passive of many degeneracy (Boogert et al.2000;Pontoppidan et al.2008). Another methodology employed byRocha et al.(2015), used the complex refractive

× E-mail: willrobson88@hotmail.com

index (CRI) of ices simulated and exposed to cosmic rays in the laboratory to fit the major and minor features due to ice processing observed in the Elias 29 circumstellar environment. The first important example of such investiagations concernes about the so-called tholins in atmosphere of Titan Khare et al.(1984). Tholins are particulate materials formed from ultraviolet processing of of gas mixtures containing mainly $N_2$ and $CH_4$ as well as other minor organic components. Other papers that present the complex refractive index of tholin are Imanaka et al.(2012);Mahjoub et al.(2012);Brassé et al.(2015).

The present paper provides a database of complex refractive index in the infrared of 13 water-bearing ices processed by cosmic rays as simulated in laboratory, indispensable inmany astronomical scenarios. Cosmic rays are ubiquitous in the interstellar medium and can reach the deepest regions not illuminated by UV and X-rays (Drury et al.2009;Cleeves et al.2013;Henning et al.2013;Cleeves et al.2014,2015).Cleeves et al.(2014) showed that cosmic rays are an important agent to ionize and process the material in the midplane in the early stage of protoplanetary disks, where the ices are formed.Boogert et al.(2008) showed that some absorption bands observed in the spectra of protostars can only be explained by assuming processing of the ices by cosmic rays. Additional discussion about the importance of cosmic rays in the interstellar chemistry can be found in the paper ofIndriolo et al.(2013). The processing of astrophysical ices by cosmic rays has been simulated in the laboratory, as shown in previous papers (Strazzulla et al.1983;Palumbo & Baratta2006;Strazzulla et al.2007;Seperuelo-Duarte et al.2009;Pilling et al.2010a; Hi-jazi et al.2011;Bergantini et al.2014;de Barros et al.2014a;Andrade et al.2014;Mejía et al.2014). These authors experimentally elucidated that cosmic rays are able to destroy chemical bonds, induce the formation of complex molecules, and lead to sputtering. All these effects lead to changes in the intrinsic properties of the ice, which are recorded in the CRI, given by the formula $\tilde{m} = n + ik$. The real part *n* determines scattering, whereas the imaginary number *k* determines absorption, when electromagnetic radiation interacts with matter.

CRI is an essential parameter in the astrophysical context, because it is employed to calculate the absorption and scattering opacity of matter. The latter is a parameter used to define how the electromagnetic waves interact with interstellar dust grains and ice mantles. Given the importance of CRI, we present in this paper the first database of CRI for several astrophysical water-containing ices bombarded by cosmic rays simulated in the laboratory at different projectile fluences (equivalent to a deposited dose). The database addresses the lack of n and k values in the literature for processed astrophysical ices. It can also be employed to understand differences in the absorption profile due to ices. Additionally, as test case, we employed the complex refractive index of an $H_2O:CO_2:NH_3:CH_4$ ice mixture as input into a radiative transfer code to study evidences of the ice processing in a typical T-Tauri model.

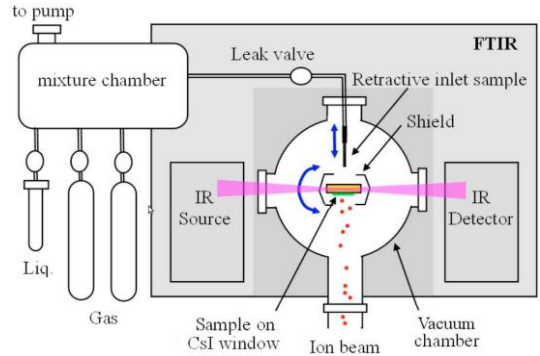

**Figure 1.** Typical experimental setup to simulate the processing of ices by heavy ion at GANIL. (Pilling et al.2010a).

This paper is organized as follows: section 2 shows the methodology to simulate the processing of the ices and how the CRI was calculated. The results and the astrophysical implications are given in Section 3 and 4, respectively. The summary is presented in Section 5.

## 2 METHODOLOGY

### 2.1 Laboratory experiments

The laboratory simulations were performed at the "Grand Accélérateur National d'Ions Lourds" (GANIL), in Caen – France by the authors of this paper in different experiments over several years. In this work, the data are compiled to build a catalogue of complex refractive indices of different astrophysical ices.

The experimental procedures are described in details inPilling et al.(2010a,b,2012);de Barros et al.(2014a). Here we briefly present the typical methodology of the experiments, as shown in Figure 1. Vapours of liquid or/and gaseous samples were mixed in a pre-chamber at pre-determined ratio and deposited by a stainless stell needle tube in the ultra-high vacuum main chamber (base pressure $<10^{-8}$ mbar) onto a cooled CsI substrate at around 13 K. The condensed molecules form an amorphous ice. After sample preparation, the ice was irradiated with heavy ions, and IR absorption spectra were recorded at different projectile fluences.

For the data compiled in this paper, two different projectiles, O and Ni, were used: $^{58}Ni^{13+}$, $^{58}Ni^{11+}$, $^{16}O^{7+}$, and $^{16}O^{5+}$. It is important to note that during the experiments the ions pass through the ice; therefore, the chemical enrichment is due to radiolysis following energy deposition by the projectiles. The projectiles themselves are not implanted and thus do not contribute to the induced chemistry (e.g. Andrade et al. (2013).

The samples were monitored in-situ by Fourier transformed infrared (FTIR) Spectrometry before irradiation and after each irradiation step at the corresponding ion fluence. The spectra were recorded from 5000 – 600 cm$^{-1}$ (2.0 - 16.6 μm). The obtained absorbance spectrum (transmittance mode) was recorded considering the following relation:

$$Abs_\nu = -\log\left(\frac{I}{I_0}\right)_\nu \quad (1)$$

where $I$ is the intensity of infrared light received by the ice and $I_0$ is the intensity transmitted by the ice at a given wavenumber ν.

Table 1 presents a list of the different water-containing ices studied in this paper. For each ice, three different spectral data were analyzed (unirradiated, low fluence and high fluence). The sample thickness, temperature, type of projectile, and fluence employed are also listed. The temperature around 13 K during the deposition simulates the cold region of interstellar medium, such as dense molecular clouds and midplane of protoplanetary disks. On the other hand, temperatures of 35 K and 72 K, are suited to simulate warm regions of protoplanetary disks. Additionally, the ions of $^{16}$O and $^{58}$Ni are able to simulate the effects of heavy cosmic-rays on the ices.

## 2.2 Thickness calculation

Thickness is a fundamental parameter to calculate the complex refractive index as shown in Section 2.3. However, the measurement of this parameter is not simple, and it may depend on instrumentation or assumptions about the physical and chemical properties of the samples.

Some papers (Baratta & Palumbo 1998; Palumbo & Baratta 2006; Hudson et al. 2014) reported a thickness measurement by using a He-Ne laser to monitor the ice thickness during de-position. By this methodology is possible to obtain an interference pattern given by interaction between light and matter during growth of the thin ice film. Despite limitations of this technique it is possible to measure the ice thickness with a precision of a few Ångström in many cases. However, such a procedure only yields the thickness of the sample during deposition of the material onto the substrate, and therefore, the unprocessed ice. The equation used to calculate the thickness, following Moore et al. (2010), is given $d = N_{fringes}\lambda/\left(2\sqrt{n^2 - \sin\theta}\right)$, in which $N_{f\,ringes}$ is the number of interference fringes observed with the laser of wavelength λ (typically 670 nm), $n$ is the refractive index of the sample measured at 670 nm and θ is the reflection angle between the incoming laser beam and the vector normal to substrate. Usually, the laser is positioned parallel 6    *Rocha et al.***Table 1.** Water-containing ices included in this paper. The first two columns give the label and the sample composition, respectively. The third and fourth column presents the thickness d of the ice and its decrease Δd/d due to processing, respectively. The subsequent columns lists the temperature of the ice, the energy and type of projectile, and the fluence (number of projectiles accumulated per surface area, i.e. the flux integrated over time) used in the experiments, respectively. The information was taken from published and yet unpublished data as indicated at the end of the table.

| Label | Sample | d (μm) | Δd/d % | Temperature (K) | Energy (MeV) - Projectile | Fluence ($10^{10}$ ions/cm$^2$) | References |
|---|---|---|---|---|---|---|---|
| D1a | $H_2O:CO_2$ (1:1) | 0.60 ± 0.05 | 0.0 ± 0.0 | 13 | ... | 0 | [1] |
| D1b | $H_2O:CO_2$ (1:1) | 0.59 ± 0.05 | 1.6 ± 0.1 | 13 | 52 - $^{58}Ni^{13+}$ | 100 | [1] |
| D1c | $H_2O:CO_2$ (1:1) | 0.49 ± 0.03 | 18.3 ± 0.2 | 13 | 52 - $^{58}Ni^{13+}$ | 1000 | [1] |
| D2a | $H_2O:CO_2$ (10:1) | 0.40 ± 0.02 | 0.0 ± 0.0 | 13 | ... | 0 | [1] |
| D2b | $H_2O:CO_2$ (10:1) | 0.39 ± 0.02 | 2.5 ± 0.2 | 13 | 52 - $^{58}Ni^{13+}$ | 100 | [1] |
| D2c | $H_2O:CO_2$ (10:1) | 0.34 ± 0.02 | 15.0 ± 1.2 | 13 | 52 - $^{58}Ni^{13+}$ | 500 | [1] |
| D3a | $H_2O:CH_4$ (1:0.6) | 2.00 ± 0.10 | 0.0 ± 0.0 | 16 | ... | 0 | [2] |
| D3b | $H_2O:CH_4$ (1:0.6) | 1.99 ± 0.10 | 0.5 ± 0.0 | 16 | 40 - $^{58}Ni^{11+}$ | 100 | [2] |
| D3c | $H_2O:CH_4$ (1:0.6) | 1.89 ± 0.09 | 5.5 ± 0.5 | 16 | 40 - $^{58}Ni^{11+}$ | 1000 | [2] |
| D4a | $H_2O:CH_4$ (10:1) | 1.00 ± 0.06 | 0.0 ± 0.0 | 16 | ... | 0 | [2] |
| D4b | $H_2O:CH_4$ (10:1) | 0.99 ± 0.06 | 1.0 ± 0.1 | 16 | 40 - $^{58}Ni^{11+}$ | 100 | [2] |
| D4c | $H_2O:CH_4$ (10:1) | 0.89 ± 0.05 | 11.0 ± 1.0 | 16 | 40 - $^{58}Ni^{11+}$ | 1000 | [2] |
| D5a | $H_2O:NH_3$ (1:0.5) | 1.40 ± 0.08 | 0.0 ± 0.0 | 13 | ... | 0 | [3] |
| D5b | $H_2O:NH_3$ (1:0.5) | 1.39 ± 0.08 | 0.7 ± 0.1 | 13 | 46 - $^{58}Ni^{13+}$ | 100 | [3] |
| D5c | $H_2O:NH_3$ (1:0.5) | 1.23 ± 0.07 | 12.1 ± 0.9 | 13 | 46 - $^{58}Ni^{13+}$ | 1600 | [3] |
| D6a | $H_2O:HCOOH$ (1:1) | 7.40 ± 0.20 | 0.0 ± 0.0 | 15 | ... | 0 | [4] |
| D6b | $H_2O:HCOOH$ (1:1) | 7.39 ± 0.20 | 0.1 ± 0.0 | 15 | 46 - $^{58}Ni^{11+}$ | 100 | [4] |
| D6c | $H_2O:HCOOH$ (1:1) | 7.29 ± 0.19 | 1.5 ± 0.1 | 15 | 46 - $^{58}Ni^{11+}$ | 1000 | [4] |
| D7a | $H_2O:CH_3OH$ (1:1) | 4.50 ± 0.12 | 0.0 ± 0.0 | 15 | ... | 0 | [5] |
| D7b | $H_2O:CH_3OH$ (1:1) | 4.49 ± 0.12 | 0.2 ± 0.0 | 15 | 40 - $^{58}Ni^{11+}$ | 100 | [5] |
| D7c | $H_2O:CH_3OH$ (1:1) | 4.39 ± 0.11 | 2.4 ± 0.2 | 15 | 40 - $^{58}Ni^{11+}$ | 1000 | [5] |
| D8a | $H_2O:NH_3:CO$ (1:0.6:0.4) | 1.70 ± 0.09 | 0.0 ± 0.0 | 13 | ... | 0 | [3] |
| D8b | $H_2O:NH_3:CO$ (1:0.6:0.4) | 1.69 ± 0.09 | 0.6 ± 0.0 | 13 | 46 - $^{58}Ni^{13+}$ | 100 | [3] |
| D8c | $H_2O:NH_3:CO$ (1:0.6:0.4) | 1.49 ± 0.08 | 12.3 ± 1.0 | 13 | 46 - $^{58}Ni^{13+}$ | 2000 | [3] |
| D9a | $H_2O:NH_3:c-C_6H_6$ (1:0.3:0.7) | 4.50 ± 0.12 | 0.0 ± 0.0 | 13 | ... | 0 | [6] |
| D9b | $H_2O:NH_3:c-C_6H_6$ (1:0.3:0.7) | 4.48 ± 0.12 | 0.4 ± 0.0 | 13 | 632 - $^{58}Ni^{24+}$ | 200 | [6] |
| D9c | $H_2O:NH_3:c-C_6H_6$ (1:0.3:0.7) | 4.18 ± 0.10 | 7.1 ± 0.6 | 13 | 632 - $^{58}Ni^{24+}$ | 3000 | [6] |
| D10a | $H_2O:CO_2:CH_4$ (10:1:1) | 2.10 ± 0.12 | 0.0 ± 0.0 | 72 | ... | 0 | [7] |
| D10b | $H_2O:CO_2:CH_4$ (10:1:1) | 2.09 ± 0.12 | 0.5 ± 0.0 | 72 | 15.7 - $^{16}O^{5+}$ | 10 | [7] |
| D10c | $H_2O:CO_2:CH_4$ (10:1:1) | 2.08 ± 0.12 | 0.9 ± 0.6 | 72 | 15.7 - $^{16}O^{5+}$ | 100 | [7] |
| D11a | $H_2O:H_2CO:CH_3OH$ (100:0.2:0.8) | 0.30 ± 0.05 | 0.0 ± 0.0 | 15 | ... | 0 | [8] |
| D11b | $H_2O:H_2CO:CH_3OH$ (100:0.2:0.8) | 0.13 ± 0.06 | 56.6 ± 4.5 | 15 | 220 - $^{16}O^{7+}$ | 1700 | [8] |
| D11c | $H_2O:H_2CO:CH_3OH$ (100:0.2:0.8) | 0.01 ± 0.07 | 96.7 ± ... | 15 | 220 - $^{16}O^{7+}$ | 9600 | [8] |
| D12a | $H_2O:NH_3:CO_2:CH_4$ (10:1:1:1) | 1.40 ± 0.05 | 0.0 ± 0.0 | 35 | ... | 0 | [9] |
| D12b | $H_2O:NH_3:CO_2:CH_4$ (10:1:1:1) | 1.39 ± 0.05 | 0.7 ± 0.0 | 35 | 15.7 - $^{16}O^{5+}$ | 10 | [9] |
| D12c | $H_2O:NH_3:CO_2:CH_4$ (10:1:1:1) | 1.38 ± 0.05 | 1.4 ± 0.1 | 35 | 15.7 - $^{16}O^{5+}$ | 100 | [9] |
| D13a | $H_2O:NH_3:CO_2:CH_4$ (10:1:1:1) | 1.10 ± 0.03 | 0.0 ± 0.0 | 72 | ... | 0 | [9] |
| D13b | $H_2O:NH_3:CO_2:CH_4$ (10:1:1:1) | 1.09 ± 0.03 | 0.9 ± 0.1 | 72 | 15.7 - $^{16}O^{5+}$ | 10 | [9] |
| D13c | $H_2O:NH_3:CO_2:CH_4$ (10:1:1:1) | 1.08 ± 0.03 | 1.8 ± 0.1 | 72 | 15.7 - $^{16}O^{5+}$ | 100 | [9] |

[1] Pilling et al.(2010b); [2] de Barros et al. (in preparation); [3] Pilling et al.(2010a); [4] Bergantini et al.(2014); [5] de Barros et al.(2014b); [6] Pilling et al.(2012); [7] Pilling et al. (in preparation); [8] de Barros et al.(2014a); [9] Bergantini et al. (in preparation).

to the inlet needle, and therefore, able to produce an interference pattern during the deposition of the gas. However, when the substrate is rotated to irradiate the ice, would be necessary another laser, positioned parallel to the flux of irradiation in order to try measure some kind of interference pattern. Additionally, the substrate should be rotated for each procedure (deposition, irradiation and spectrum collect), and for each case, the angle should be exactly the same otherwise, any difference produces error in the thickness calculation.

On the other hand, many other papers (Pilling et al.2010a,b; Bergantini et al.2014; de Barros

MNRAS **000**, 1–23 (2015)

et al.2014a,b;Lv et al.2013;Dartois et al.2013), have shown that is possible to calculate the thickness of unprocessed ices from the absorbance spectra by using the following equations:

$$d(\mu m) = \left[\frac{N}{\rho}\frac{\mathcal{M}}{\mathcal{N}_A}\right] \times 10^4 \qquad (2a)$$

$$N = \frac{2.3}{A} \int_{\nu_1}^{\nu_2} Abs_\nu d\nu \qquad (2b)$$

where N is the column density (molecule cm$^{-2}$), specific density (g cm$^{-3}$), molar mass, the Avogrados's number, A the band strength of the of molecular bond and Abs$_\nu$ the absorbance measured at laboratory.

Equations (2a) and (2b) were employed to calculate the thickness of the unprocessed samples, for the fluence 0 ions cm$^{-2}$, showed in the Table 1. Errors are due to uncertainties from the band strength and density values used. The band strength employed by authors, are often obtained from the literature, even though those values were estimated for pure ice (e.g. $H_2O$, $CO$, $CH_3OH$), as can be seen inHudgins et al.(1993) andBouilloud et al.(2015). In our case, on the other hand, to calculate the thickness of virgin ices, we used the band strength of the O-H stretching mode of $H_2O$ corresponding to "weak mixture". This term was introduced byHudgins et al.(1993) which means a water ice in a matrix containing polar species with low abundance.In this case, we adopted the band strength of the O-H stretching mode at 3298 cm$^{-1}$ (3.03 µm) as being equal to 2.1 $\times$ 10$^{-16}$ cm/molecule. However, this issue is difficult to address, because would be necessary the knowledge of all charge distributions interacting with each other inside the ice would be necessary. Processing of ices by energetic ions induces many processes such as molecular dissociation and formation, sputtering and sample compaction. Sputtering (ejection) of molecules leads the erosion of a surface and it is the main agent that lead to decreases in the thickness of ices (e.g. Baragiola et al.(2003)). The number of molecules ejected after ion impact (sputtering yield – Y, in molecules ion$^{-1}$) was found to scale with the square of the electronic projectile energy loss (Seperuelo-Duarte et al.2010;Dartois et al.2015;Mejía et al.2015). For experiments at GANIL with heavy ions of energy about 50 MeV, yields as high as Y = 10$^5$ molecules ion$^{-1}$ were observed. Assuming that each monolayer of ice-matrix has 1 nm of thickness, as estimated byPilling et al. (2012), we can calculate the final thickness of ices after each experiment as function of the fluence. The numbers are shown in Table 1. The main error source in the thickness of the processed ice is coming from the sputtering yield, which leads to a maximum error around 9 %. Additionally, compaction is observed during the experiments, as shown by Palumbo & Baratta (2006) and in the case of heavy ions by Dartois et al. (2013). Such effects are seen both in laboratory spectra Pilling et al. (2010a,b), as wll in as observational ones (Boogert et al. 2008; Zasowski et al. 2009). The effect is small compared with the sputtering process, and we are considering it negligible.

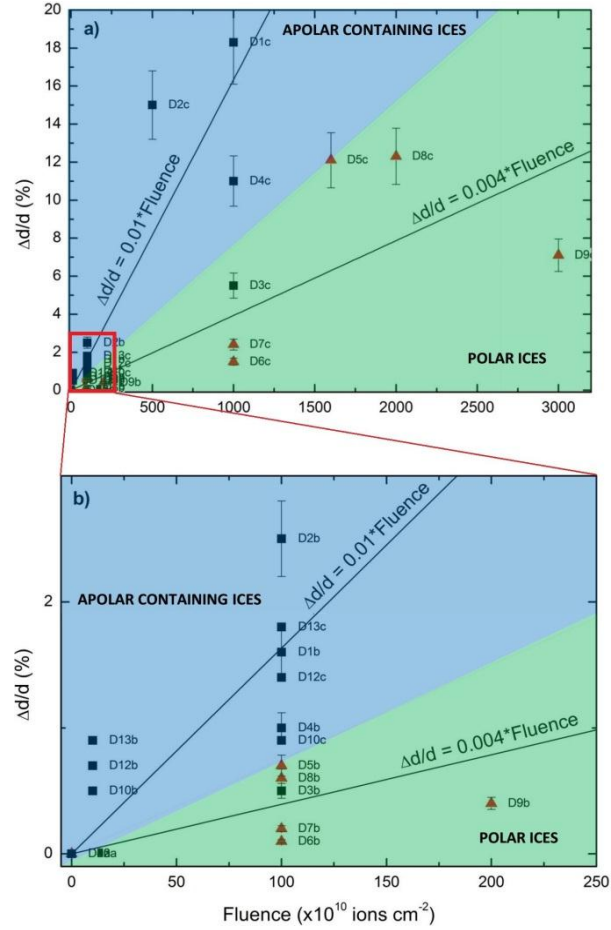

**Figure 2.** Variation in the ice thickness as function of the fluence. Black lines are fitting for polar and apolar containing ices. (a) High fluences (up to $3 \times 10^{13}$ ions cm$^{-2}$). Red box limits the regions shown in the panel (b) at low fluences (up to $2.5 \times 10^{12}$ ions cm$^{-2}$).

Figure 2a and 2b show the variation of $\Delta d/d$ as function of the fluence, between $0 - 3 \times 10^{13}$ ions cm$^{-2}$ and $0 - 2.5 \times 10^{12}$ ions cm$^{-2}$, respectively. The colors limit the regions for polar ices (green) and apolar containing ices (blue). Black lines are linear fitting for both regions. By comparing these two regions it becomes evident that the decrease in the thickness of ice depends also on the degree of polarity of the molecules, because the strength between the bonds in polar molecules is larger than bonds in polar-apolar ones. Indeed, bonds between polar molecules occur due to permanent dipole momenta whereas between polar and apolar ones they are due to induced dipole moments. In this Figure, however, the case D3 ($H_2O:CH_4$ - 1:0.4) was an exception. For all molecules, the fitting is valid for low fluences (up to $2.5 \times 10^{12}$ ions cm$^{-2}$) and for high fluences (up to $3 \times 10^{13}$ ions cm$^{-2}$).

Polar and apolar phase of astrophysical ices can be found in the interstellar medium. As pointed out byBoogert et al.(2015), polar ices exists in regions with a high H /CO gas ratio, T > 20 K, n ≥ $10^3$ cm$^{-3}$ and visual extinction $A_V$ > 1.5 mag whereas apolar ices are formed onto polar ices in regions with small H/CO gas ratio, T < 20 K, n ≥ $10^4$ cm$^{-3}$ and visual extinction $A_V$ > 3 mag. From Figure 2, is possible to conclude that rich chemistry induced by cosmic rays inside molecular clouds exists, partly because cosmic rays break bonds in the apolar phase of ices with more efficiency than in the polar phase.

## 2.3 Complex refractive index calculation

To calculate the CRI, the computational code NKABS ("Determination of N and K from AB-Sorbance data") from Rocha et al.(2014) was applied. This code calculates the real and imaginary part of the refractive index directly from absorbance data obtained in the laboratory. It was developed in the Python programming language, and it is available at the following web site: http://www1.univap.br/gaa/nkabs-database/data.htm. Briefly, the code is divided into four steps: i) the imaginary part *k* is calculated by using the Lambert-Beer law, given by:

$$\alpha = \frac{1}{d}\left[2.3 \cdot Abs_v + ln\left|\frac{\tilde{t}_{01}\tilde{t}_{12}/\tilde{t}_{02}}{1+\tilde{r}_{01}\tilde{r}_{12}e^{2i\tilde{x}}}\right|^2\right], \quad [cm^{-1}] \quad (3a)$$

$$\alpha = 4\pi v k, \quad [cm^{-1}] \quad (3b)$$

where α is called absorption coefficient, *d* is the thickness of the ice in units of centimeter; $Abs_v$ is the absorbance in infrared; $\tilde{t}_{01}, \tilde{t}_{12}, \tilde{t}_{02}, \tilde{r}_{01}$, and $\tilde{r}_{12}$ are complex coefficients of transmission and reflection, also called Fresnel's coefficients; and $\tilde{x} = 2\pi v d \tilde{m}$ in which $v$ is the wavenumber in units of cm$^{-1}$ and $\tilde{m}$ is the complex refractive index. In this step the code calculates first the Equation 3a and after that the value of *k* using Equation 3b. ii) Calculation of the real part *n* of the refractive index employing Kramers-Kroning relations is given by:

$$n(v) = n_0 + \frac{2}{\pi}\wp\int_{v_1}^{v_2}\frac{v'k(v')}{v'^2 - v^2}dv' \quad (4)$$

where $n_0$ is the refractive index at 670 nm, fromHudgins et al.(1993), and $\wp$ is the Cauchy principal value. The numerical calculation employs the Maclaurin methodOhta et al.(1998). iii) The next step is the calculation of the theoretical IR spectrum and its comparison with a measured spectrum obtained in the laboratory. iv) The best values of *n* and *k* are obtained by analyzing the error during the comparisons between theoretical and laboratory spectra. This is performed by using the Mean Average Percentage Error (MAPE). If the value is acceptable by the user, the code stops and shows the results. Otherwise, the code runs again, but improves the Fresnel's coefficients, by using the last values calculated for *n* and *k*.

The main source of error in the values of *n* and *k* are coming from the measurement of the absorbance area (*A*) of specific bands in the infrared spectra and sample thickness (*d*) because of the reasons discussed in Section 2.2. In this paper, the error was evaluated by using the method described in del Pozo & Diaz (1992), and the it was found to be around 12 %. for n and k.

## 3 RESULTS

We present the CRI of 13 water-containing ices calculated with the NKABS code in Figures 3 - 5. The results are available as ASCII files at the website: http://www1.univap.br/gaa/nkabs-database/data.htm. Each panel contains the real and imaginary part at different wavenumbers in the infrared (5000 – 600 cm$^{-1}$), corresponding to 2.0 - 16.6 µm. The figures show that the refrac-tive index changes with the projectile fluence during irradiation. These changes also depend on the initial ice composition and the fluence, and they are attributed to the emergence of new species in the sample, the destruction of parental species, as well as morphological changes of the sample. The fluences used for each experiment are shown in the panels, in which the black line indicates the unprocessed ice spectrum, the blue line an ice spectrum at intermediary fluence, and the red line an ice spectrum at the final fluence.

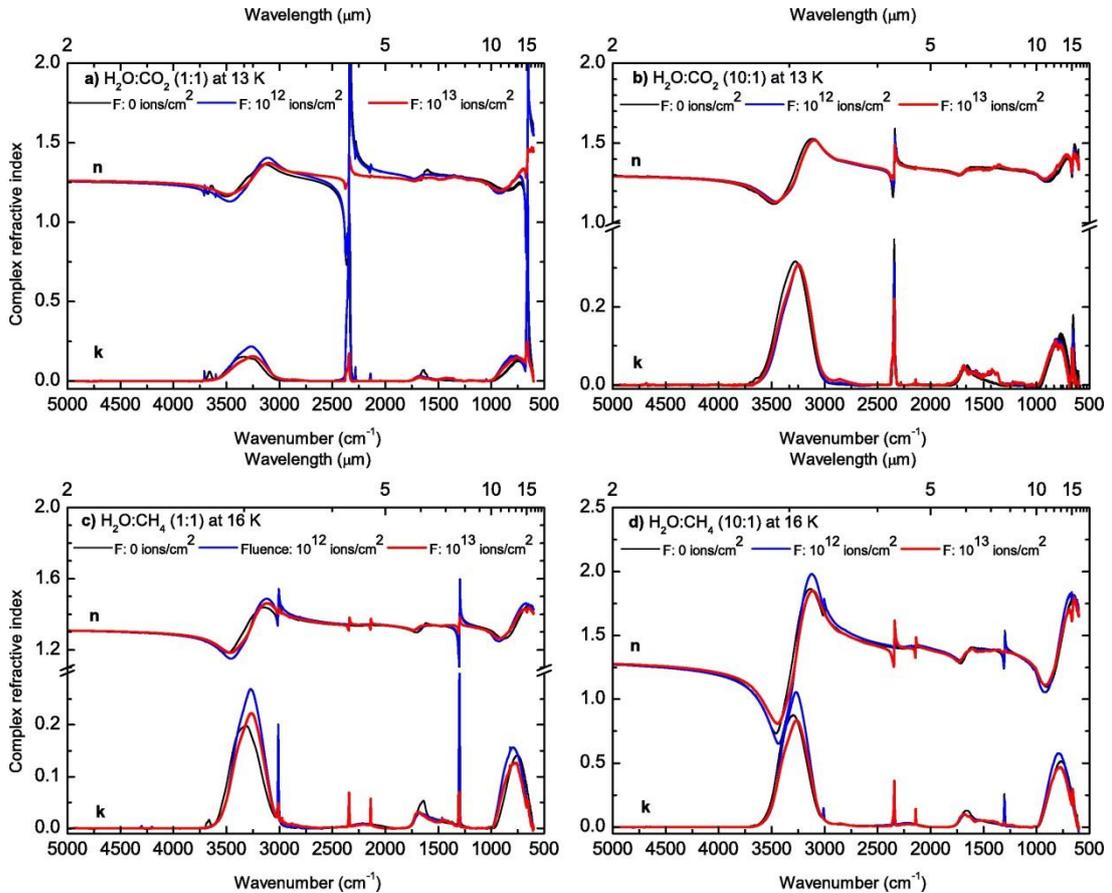

**Figure 3.** Complex refractive index of $H_2O:CO_2$ (1:1) at 13 K; $H_2O:CO_2$ (10:1) at 13K; $H_2O:CH_4$ (1:1) at 16K; and $H_2O:CH_4$ (10:1) at 16 K at different ion fluences (see Table 1).

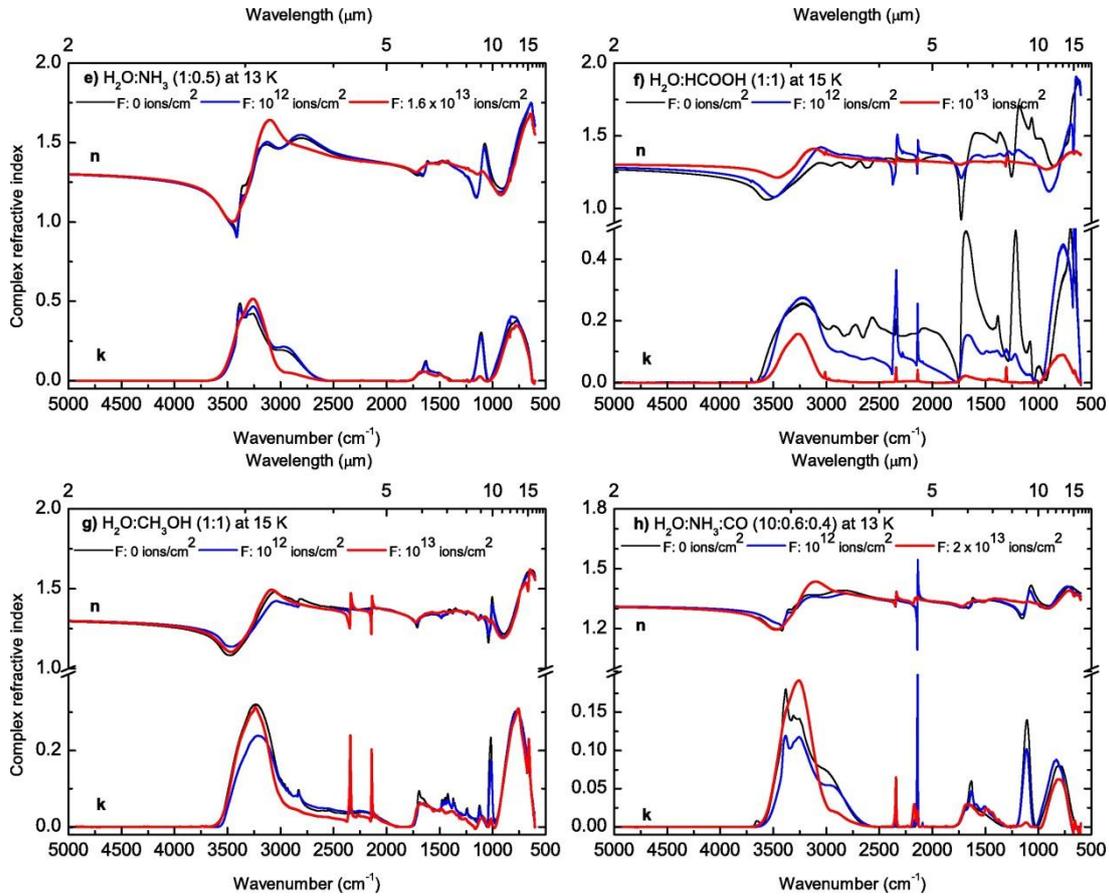

**Figure 4.** Complex refractive index of $H_2O:NH_3$ (1:0.5) at 13 K; $H_2O:HCOOH$ (1:1) at 15K; $H_2O:CH_3OH$ (1:1) at 15K; and $H_2O:NH_3:CO$ (10:0.6:0.4) at 13 K at different ion fluences (see Table 1).

The destruction of parental species of the ice and the formation of new molecules is derived from the variation in the molecular column density at different fluences as deduced from the measured FTIR spectra. Details about the newly formed species in the experiments with the ices showed in Table 1 can be found in the respective indicated publications. Briefly, however, as examples we note the formation of carbonic acid ($H_2CO_3$) during the experiments with $H_2O:CO_2$ (see D1 label in Table 1). This species was found in several ion bombardment experiments (e.g. Moore & Khanna(1991);Gerakines et al.(2000)). The dissociation of $H_2O$ molecules by the projectiles is followed by electron attachment to $CO_2$ or OH, forming $HCO^-$. Finally, this latter ion reacts with a proton, producing $H_2CO_3$. This molecule has important astrophysical implications as far surfaces of moons such as Europa, Calisto and Iapetus are concerned, because they evolve within the magnetospheres of their host planets (Peeters et al.2008) and are subjected to high fluxes of bombarding ions. On the other hand, another kind of techniques such as sen- sitive two-dimensional gas chromatography time-of-flight mass spectrometry (GC×GC-TOFMS) may be used to observe the formation of new molecular species from ice processing. For instance, Meinert et al.(2016) observed the formation of prebiotic species such as ribose and related sugars, important from the astrobiological point of view, from the UV processing of a simulated ice composed by $H_2O$, $CH_3OH$ and $NH_3$.

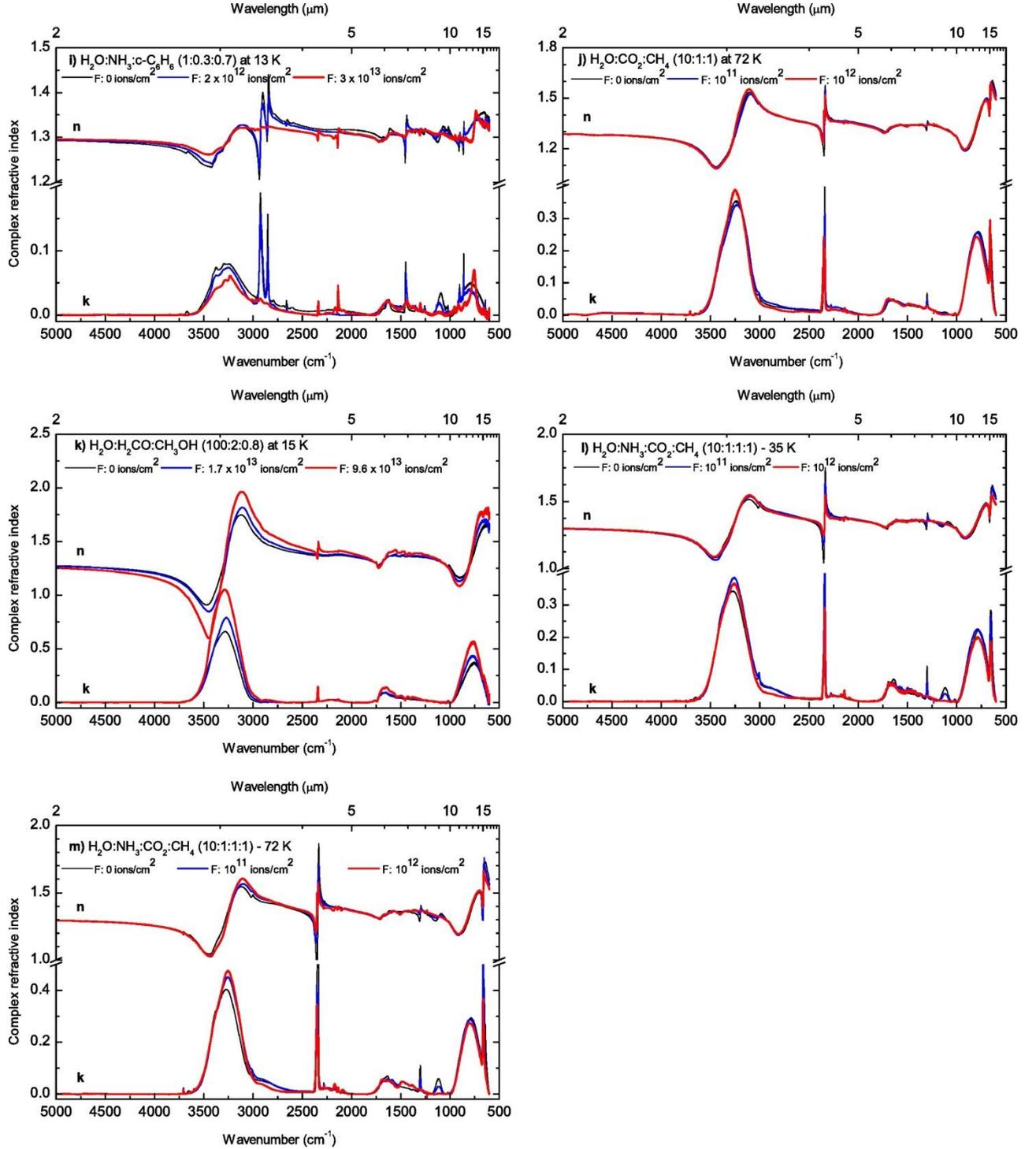

**Figure 5.** Complex refractive index of $H_2O:NH_3:c-C_6H_6$ (1:0.3:0.7) at 13 K; $H_2O:CO_2:CH_4$ (10:1:1) at 72 K; $H_2O:H_2CO:CH_3OH$ (100:2:0.8) at 15 K; $H_2O:NH_3:CO_2:CH_4$ (10:1:1:1) at 35 K; and $H_2O:NH_3:CO_2:CH_4$ (10:1:1:1) at 72 K at different ion fluences (see Table 1).

Although details about the experimental procedure to simulate the ices presented in this paper are already written in their respective references (see Table 1), some general features about the ices should be highlighted for Figures 3 - 5. For instance, the OH band at 3.03 μm shows that the chemical environment, leading to differences in the peak position and shape of the band, has strong influence on the molecular species in the ice. Furthermore, the band intensity increases with the fluence, meaning that energetically processing of the ice is producing more $H_2O$ (e.g. Figure 3a, 3c, 3d, 4h, 5k). Some molecules, on the other hand, are completely destroyed, for example HCOOH in a matrix with $H_2O$ (Figure 4f). As pointed out by Bergantini et al.(2014), the destruction cross section of formic acid in the presence of water is larger than that of pure HCOOH. Indeed, as can be seen in Figure 4f, the ice at the final fluence is composed mainly by $H_2O$, CO and $CO_2$. However, the most striking evidence of the ice processing can be seen between 1750 - 1250 $cm^{-1}$ (5.7 - 8.0 μm): there is evidence of formation of many organic molecules such as HCOOH, $CH_3OH$, $CH_3CHO$ and $H_2CO$. This features highlights how important the chemistry in the condensed phase is.

Table 2 presents some parameters considered in the models after the calculation of the refractive index. Note, however that the value for $n_0$ was determined by a weighted mean of values measured separately for the species in the ice composition. The low MAPE means that NKABS obtained indeed good values of the n and k to reproduce the absorbance spectrum obtained in the laboratory.

**Table 2.** The values for $n_0$ and MAPE obtained from the NKABS code.

| Parameters | D1 | D2 | D3 | D4 | D5 | D6 | D7 | D8 | D9 | D10 | D11 | D12 | D13 |
|---|---|---|---|---|---|---|---|---|---|---|---|---|---|
| $n_0$ [a,b,c] | 1.27 | 1.30 | 1.30 | 1.32 | 1.33 | 1.32 | 1.31 | 1.32 | 1.31 | 1.32 | 1.32 | 1.32 | 1.32 |
| MAPE[a] ($\times 10^{-4}$ %) | 0.8 | 1.8 | 0.3 | 0.2 | 0.2 | 13.7 | 5.8 | 2.1 | 5.1 | 5.1 | 2.7 | 3.4 | 3.3 |
| MAPE[b] ($\times 10^{-4}$ %) | 1.1 | 0.7 | 0.4 | 0.3 | 0.3 | 10.6 | 5.1 | 1.7 | 4.5 | 5.0 | 1.7 | 3.6 | 4.0 |
| MAPE[c] ($\times 10^{-4}$ %) | 0.6 | 0.5 | 0.3 | 0.2 | 0.3 | 7.6 | 5.3 | 2.3 | 4.2 | 5.2 | 5.5 | 3.3 | 4.1 |

[a] unirradiated sample
[b] processed sample
[b] highly processed sample

# 4 ASTROPHYSICAL IMPLICATIONS

As already discussed throughout this paper, cosmic rays play an important role in the chemistry of cold and dense regions inside astrophysical environments because they can penetrate deeper in molecular clouds and circumstellar environments triggering chemical modification also in places which stellar photons cannot reach. Inside of these regions, the temperature is not high enough to allow chemical reactions such as neutral-neutral chemistry (Henning et al.2013). In such cold regions, only cosmic rays can ionize efficiently and consequently drive a rich chemistry by many ways such as ion-molecule collisions, charge transfer and dissociative recombination (Tielens 2005). In view of the observations of ice processing in several protostars, let us now implant processed ices into a generic T-Tauri model.

The cosmic ray energy spectrum (Cronin et al.1997) measured in the Solar System shows that the flux of particles decreases with their kinetic energy. For instance, that spectrum between $10^{10}$ eV and $10^{21}$ eV can be weel described by a power law ($\phi \propto E^{-2.7}$ -Indriolo et al.(2013)). For energies below 1 MeV, the flux of cosmic rays has been measured by Voyager and Pioneer farther away from the zone strongly influenced by the solar wind. Shen et al.(2004) report the flux of protons with energy of 16 MeV to be about 0.2 cm$^{-2}$ s$^{-1}$, whereas the flux of heavy ions such Oxygen (16 MeV) is around $5 \times 10^{-4}$ cm$^{-2}$ s$^{-1}$. On the other hand, measurements obtained by Mewaldt et al.(2007) show that the flux of 16 MeV oxygen ions is around $1 \times 10^{-2}$ cm$^{-2}$ s$^{-1}$ in the Solar System. The scenario for protoplanetary disks is different, because the magnetic field coupled to stellar wind modulates cosmic rays with energy below 1 GeV (Cleeves et al. 2013). According to Cleeves et al.(2013,2015) the flux of particles with energy of 16 MeV in the protostellar disk is around $2 \times 10^{-7}$ cm$^{-2}$ s$^{-1}$. In the experiments, however, the projectile fluxes were around $10^{9}$ cm$^{-2}$ s$^{-1}$ to be able to perform the experiments in a reasonable time. The fluences employed in the laboratory and observed in interstellar medium are of course similar. For instance, assuming that flux of particles for Class I protostars ($\sim 10^5$ years) corresponds approximately to the flux of Oxygen estimated by Mewaldt et al.(2007), the fluence is around $3 \times 10^{10}$ ions cm$^{-2}$. For Class II protostars ($\sim 10^6$ years), for example, the fluence is around $7 \times 10^6$ ions cm$^{-2}$. Hence, it is possible that the part of the chemistry observed in the ices for protostellar disks (class II), was driven by cosmic rays in the previous stage (e.g. class I phase).

**4.1 T-Tauri model**

In this paper, as a test case, the CRI virgin and processed ice containing $H_2O:NH_3:CO_2:CH_4$ (see Figure 5l) was employed in a generic model of T-Tauri disk, as described in Whitney et al. (2003a,b), to simulate an object in the early class II phase. The mixture used in this paper is important for Young Stellar Objects because it contains the most abundant molecular species observed toward star-forming regions (Öberg et al. 2011). Additionally, the fluence used for this ice was between $10^{10}$ - $10^{12}$ ions cm$^{-2}$, the values agree with the typical fluence expected for protoplanetary disks. The model central star has a mass of 1 M , radius of 2 R and effective temperature of 4000 K. In this model it is assumed that this star emits a pure blackbody spectrum. The accompanying disk of this star has a radius of 200 AU and mass of 0.02 M . The inner radius of disk is given by relation $R_{in}$ = R ($T_{silicate}$/T ) as showed by Dullemond & Monnier (2012). Assuming a sublimation temperature of silicate of 1500 K (Gail 2010), the inner radius of the disk is 0.06 UA. Additionally the disk is in hydrostatic equilibrium in the vertical direction and its density profile $\rho(r, \theta)$ is given by the equation from Dullemond (2000):

$$\rho(r,\theta) = \frac{\Sigma_0 (r/R_0)^{-1}}{\sqrt{2\pi} H(r)} exp \left\{ -\frac{1}{2} \left[ \frac{r \cos \theta}{H(r)} \right]^2 \right\} \quad (5)$$

$r^-$

where in the Equation 5 $\Sigma_0$ is the superficial density at outer edge of the disk, $R_0$ the external radius of the disk, and $H(r)$ is the disk scale height. The last variable is defined from the self-irrariated passive disk proposed by Chiang & Goldreich (1997) and is given by H(r) = r · $H_0/R_0$ · $(r/R_0)^{2/7}$. The ratio $H_0/R_0$ < 0.1 defines cold disks whereas $H_0/R_0$ > 0.1 hot disks according to Faure et al. (2014). Additionally, its typical values for protoplanetary disks ranges from 0.1 - 0.3 (Ono et al. 2014), and hence for a generic case, was adopted the value of 0.17.

**4.2 Dust and ice model**

Since the interaction between light and matter in astrophysical environments is regulated by absorption and scattering processes, we calculated the opacity for a core of silicate[1] (Weingartner & Draine 2001) with inclusion of amorphous carbon and ice (H $_2$O:NH$_3$:CO$_2$:CH$_4$) to simulate the presence of a mantle. The carbon inclusion was assumed to be 0.15 (Pontoppidan et al. 2005b), and the abundances of ice relative to H$_2$ were $10^{-4}$ for water and $10^{-5}$ for CO$_2$, CH$_4$ and NH$_3$. To perform the mixing rule we adopted the Maxwell-Garnett effective medium theory (Bohren & Huffman 1983) to calculate the e ffective complex refractive index for the grains. Considering that relation between complex dielectric function ($\tilde{z}$) and complex refractive index ($\tilde{m}$) for non-magnetic materials is given by $\tilde{z} = \tilde{m}$, the equation for effective complex refractive index ($\tilde{m}_{av}$) is given by:

$$\tilde{m}_{av}^2 = \tilde{m}_{mat}^2 \left[ 1 + \frac{3f(\tilde{m}_{inc}^2 - \tilde{m}_{mat}^2)/(\tilde{m}_{inc}^2 + 2\tilde{m}_{mat}^2)}{1 - f(\tilde{m}_{inc}^2 - \tilde{m}_{mat}^2)/(\tilde{m}_{inc}^2 + 2\tilde{m}_{mat}^2)} \right] \quad (6)$$

where $\tilde{m}_{mat}$ and $\tilde{m}_{inc}$ are the CRI for the silicates and for inclusions, which can be carbonaceous material or ices, respectively. The opacity for the grains was calculated employing the Mie theory (Mie 1908), assuming spherical grains with a single grains size of 0.3 µm. Figure 6a – 6c show the opacities for the grains considering the Fluence 0, Fluence I ($10^{11}$ ions cm$^{-2}$) and Fluence II ($10^{12}$ ions cm$^{-2}$).

---

[1] https://www.astro.princeton.edu/ draine/dust/dustmix.html

As can be seen in Figure 6, the CRI presented in this paper allows to calculate an opacity database to be used in astrophysical models. This kind of database has been made by Ossenkopf et al. (1994), although such data are consistent only for non-processed regions in the interstellar medium. However, the processing of the ice is a common scenario and the opacities from Ossenkopf et al. (1994) lead to wrong conclusions about the chemical evolution of the ice. In this way, the database presented in this paper may be used to calculate new opacity tables, by using realistic data of processed ices, as well as realistic grain geometry.

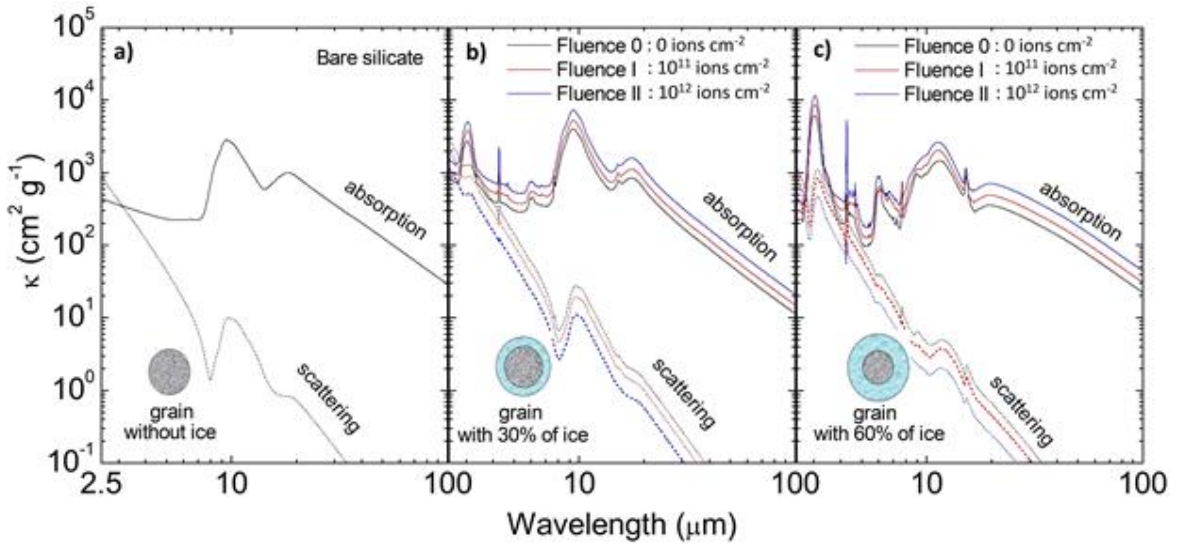

**Figure 6.** Absorption and scattering opacities of astrophysical grains employed in this paper. Panel (a) present the opacities of bare silicate and panels (b) and (c) opacities of silicate with ice at fluences 0 - II. The lines were moved arbitrarily to highlight differences due to energetic processing.

In T-Tauri model, the position of the grains was defined by their sublimation temperature. Bare silicates were employed in all disk with temperatures below 1500 K (Gail 2010) whereas grains with inclusions of ice were fixed at regions with temperatures below 50 K. In spite the sublimation temperature of $H_2O$ to be around 150 K, the ice employed in this paper was produced at around 35 K.

### 4.3 Optical depth and comparisons

Different observational spectra of protostars were taken from literature in order to calculate their optical depth in the infrared as shown in Figures 7 and 8. Such spectra were obtained from c2d Legacy Program (Evans et al.2003), covering 5.3 – 40 μm. This program was performed with the IRS instrument onboard Spitzer Space Telescope by using "short-low" and "long-low" modules (SL and LL; R = $\lambda/\Delta\lambda$ = 60 – 120, respectively). Information about these YSOs are summarized in Table 3, but more information can be found in Boogert et al.(2008). Disk inclinations of these YSOs are 65°, 69° for Elias 29 and CRBR 2422.8 – 3423, respectively. However, disk inclination for EC 90, EC 92 and IRAS 13546-3941 has not been yet been determined, due to high visual extinction toward them.Ioppolo et al.(2013) shows that they are embedded in clouds with extinction of around $A_V \sim$ 20 mag.

**Table 3.** Observational YSOs used for comparison with a T-Tauri model. Column 1 shows the name of the source and columns 2 and 3 its right ascension and declination. Column 4 the host cloud, and columns 5 and 6 their spectral index in the infrared and classification, respectively.

| Source | R.A. (J2000) | DEC (J2000) | Cloud | $\alpha_{IR}$ | Class[c] |
|---|---|---|---|---|---|
| EC 90[a] | 18h29'57.75" | +01° 14'05.9" | Serpens | -0.09 | Class I/II |
| EC 92[a] | 18h29'57.88" | +01° 12'51.6" | Serpens | +0.91 | Class I |
| IRAS 13546 - 3941[a] | 13h57'38.94" | -39° 56'00.2" | BHR 92 | -0.06 | Class I/II |
| CRBR 2422.8-3423[a] | 16h27'24.61" | -24° 41'03.3" | Ophiuchi | +1.36 | Class I |
| Elias 29[b] | 16h27'09.42" | -24° 37'21.1" | Ophiuchi | +0.53 | Class I[d] |

[a] Boogert et al.(2008).
[b] Spectrum take from ISO (Infrared Space Observatory observations – de Graauw et al.(1996) andClegg et al.(1996))
[c] Based on Lada's classification (Lada et al.1987).
[d] Spectral index enhanced due to foreground extinction

Two synthetic spectra of a typical T-Tauri model where simulated with the RADMC-3D code (Dullemond e al. *in prep.*) in order to compare their optical depth with those found for observed spectra as listed in Table 3. One of the models employed opacity with 30 % of ice, whereas the other employed opacity with 60 % of ice at two different fluences. A disk with inclination of 70° was adopted in the model because it is representative for Elias 29, CRBR 2422.8 - 3423, and EC 82. Since the sources EC 90, EC 92 and IRAS 13546-3941 are highly embedded, is probable that ices observed toward these sources are placed in the foreground clouds. Due to that, the disk inclination does not matter, and the angle of 70° can be safelyemployed.

Figure 7 shows the comparison among optical depth of the T-Tauri model by employing grains with 30 % of virgin and processed ice and the YSOs of Table 3. Figures 7a1 - 7c1 show the wavelength between 5.5 - 20 µm indicating the presence of bending mode of $H_2O$ at 6.0 µm (1666 cm$^{-1}$) and several organic compounds, silicate silicate at 9.8 µm due to Si - O vibration mode, $H_2O$ bending mode at 12.5 µm (800 cm$^{-1}$) and $CO_2$ bending mode at 15.2 µm (667 cm$^{-1}$). It is worth noting that the libration band of $H_2O$ lies at around 11.5 - 13.6 µm because its center and width are sensitive to ice composition (Hagen et al.1993). In the model, the abundance of the silicates is larger than observed for Elias 29 (blue dot) and CRBR 2422.8-3423 (red dot), which means that these objects should have a mass smaller than 0.02 M . Indeed, as shown in Pontoppidan & Dullemond (2005a) and Rocha et al. (2015), the disk masses associated with Elias 29 and CRBR 2422.8-3423 are, respectively, 0.003 M and 0.0015 M . On the other hand, the feature due to silicates is absent in EC 90, EC 92, IRAS 13546-3491. The reason for that is associated with the abundance of ice relative to silicate grain. If the grains contain around 60 % of ice in their total volume, the absorption feature due to silicate is hidden (see Figure 8). A similar conclusion was obtained by Ossenkopf et al. (1994), by using ices in which 45 % of the volume was ice. In this case, only the librational mode of water can be seen. Another important characteristic observed in Figure 7a1 - 7c1 is that CRBR 2422.8-3423 does not present the $H_2O$ ice librational band, whereas Elias 29 does. So far, is not clear why the absorption due to the water librational band is present in some protostars, and in others it is not. However, laboratory experiments performed by Öberg et al. (2007), show that the libration mode of $H_2O$ is highly sensitive to abundance of $CO_2$ (apolar molecule) and temperature. The band strength of the water libration mode decreases with abundance of $CO_2$, whereas it increases with temperature. Therefore, is possible that the missing $H_2O$ libration mode in CRBR 2422.8-3423 is associated with the abundance of $CO_2$ in the ice mantle toward this object.

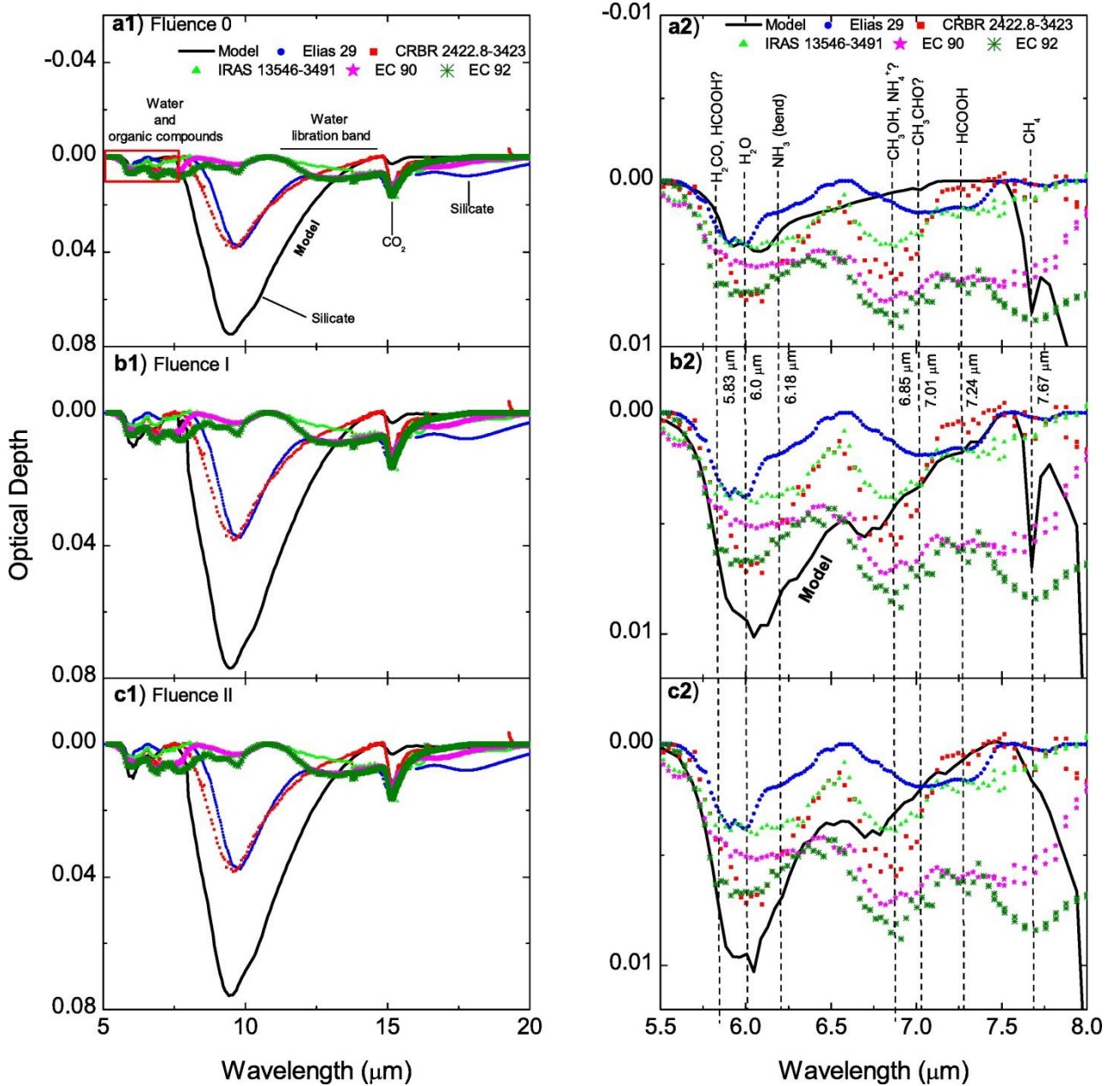

**Figure 7.** Optical depth of T-Tauri model with grains containing 30 % of ice and disk angle of 70° compared to optical depth of observed spectra. Panel (a1) show the regime between 5.0 - 20.0 µm and the red box indicate a zoom in between 5.5 - 8.0 µm showed by the panel (a2). That is the same for panels (b1, b2) and (c1, c2).

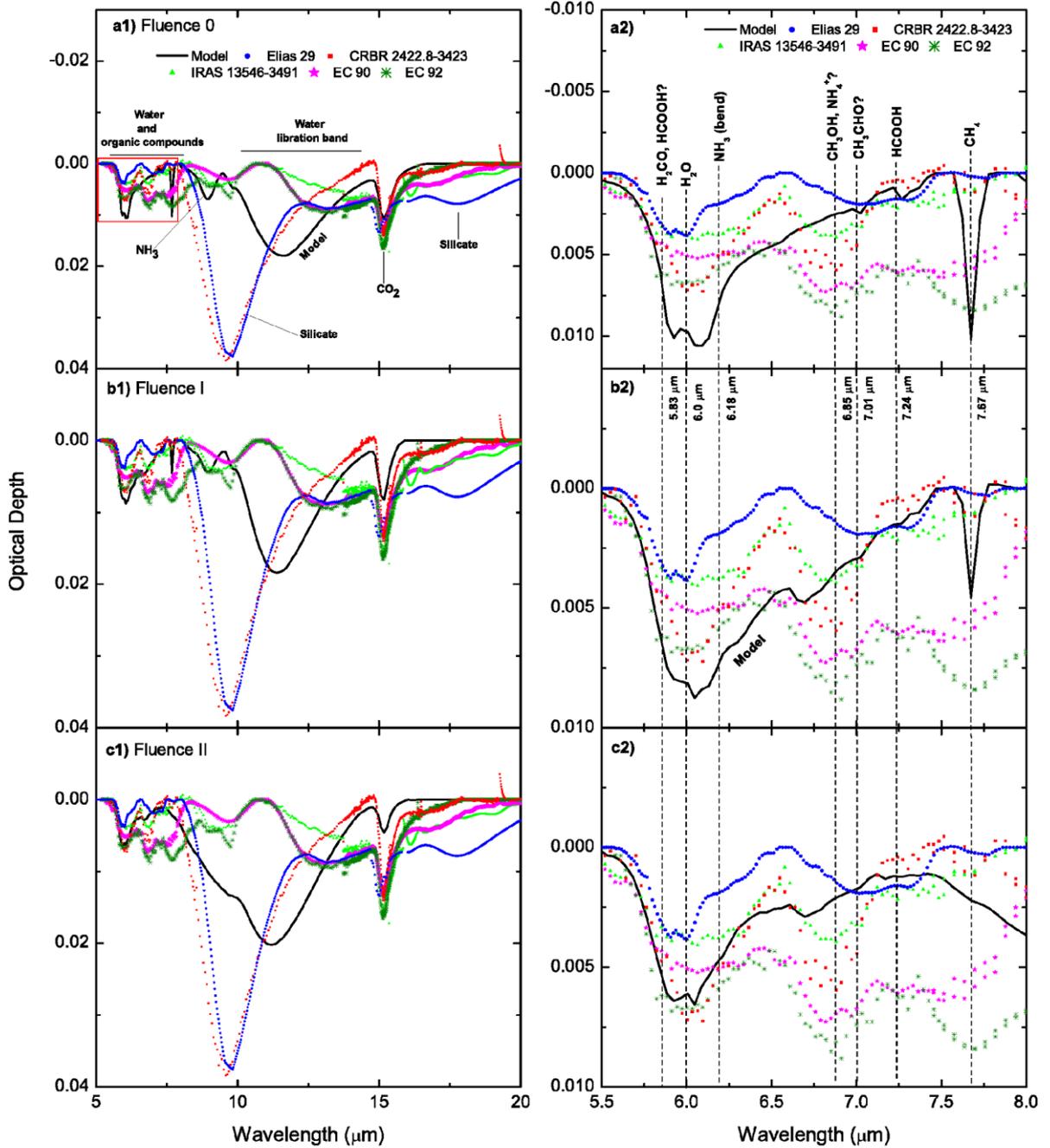

**Figure 8.** Optical depth of T-Tauri model with grains containing 60 % of ice and disk angle of 70° compared to optical depth of observed spectra. (a1) - (c1) Regime between 5.0 – 20.0 μm. Red box indicate the panels (a2) - (c2), which are a zoom in between 5.5 – 8.0 μm.

Figure 7a2 - 7c2 present a zoom in between 5.5 – 8.0 µm of panels a1 - c1 as indicated by the red box. This region is characteristic of plenty of organic compounds such as $H_2CO$, $HCOOH$, $CH_3OH$, $CH_3CHO$ and $CH_4$. By comparing the model with 30 % of virgin and processed ice with the observed protostars, we observe that virgin ice (panel a2) is able to reproduce only the features of the $H_2O$ bending mode at 6.0 µm (1666 cm$^{-1}$) and the $CH_4$ deformation mode at 7.67 µm (1303 cm$^{-1}$). On the other hand, processed ices with fluences I and II (panels b2 and c2) reproduce the region between 6.5 - 8.0 µm much better. However, the feature due to $CH_4$ cannot be fitted by the model at Fluence II, which probably means that the flux of cosmic rays that can drive the chemistry is not high, but rather moderate (Fluence I model). In order to reproduce all these spectra from protostars better, two ways come into mind: (i) perform more laboratory experiments employing the same mixture as in this paper, but by using different proportions of molecules; (ii) using other ices found in Table 1.

Figure 8 shows the comparison among optical depth of the T-Tauri model by employing grains with 60 % of virgin and processed ice and the YSOs of Table 3. Figures 8a1 - 8c1 show the same features as 7a1 - 7c1, but the absorption due to molecules in the ice is much more evident. Besides features of water and organic compounds, the presence of the umbrella vibration mode of $NH_3$ at 9.01 µm (1109 cm$^{-1}$) can be seen, as well as its energetic processing. In panels 8a1 - 8c1, the $NH_3$ molecule is destroyed by cosmic rays, and Fluence I can reproduce its profile better. In contrast to Figure 7, the volume of ice employed in this model is able to hidden the profile due to Si - O stretching mode of silicate. In this case, the model employing 60 % of ice in the grain is not representative for Elias 29 and CRBR 2422.8-3423, but only for EC 90, EC 92 and IRAS 13546-3491. The accretion rate of species onto dust grains is related with dust cross section and sticking coefficient, the later is associated with the temperature. The sticking coefficient is usually adopted as being 1 at very low temperatures, but this value can change for temperatures above 10 K (Cuppen et al.2013). Moreover, as pointed out by Boogert et al.(2015), is expected that mantle growth should be more important for silicate grains smaller than 0.1 µm, than for large grains. The size of grains in the interstellar medium has been assumed to follow a power-law (n(a) $\propto$ a$^{-3/5}$) called MRN distribution (Mathis et al.1977). According to this approximation, the graphite grains range in size between 0.005 and 1.0 µm, whereas silicate grains range from 0.025 to 0.25 µm. The lower size cut-off for silicate grains is still under debate, values between 0.0003 - 0.04 µm were reported for different directions in the interstellar medium. Moreover, results from Clayton et al.(2003) found good fits for the interstellar extinction by using the Maximum Entropy Method (MEM), yielding silicate sizes between 0.125 - 3.0 µm toward our Galaxy and Magellanic Clouds. In this way, we can infer that the grain size toward EC 90, EC 92 and IRAS 13546-3491 should follow a distribution with a peak around 0.1 µm to allow the maximum growth of the ice mantle. Additionally, it is important to note the displacement between the libration mode of $H_2O$ of model and observed protostars. As discussed above, the peak position of this band is sensitive to the ice composition.

Figure 8a2 - 8c2 present a zoom in between 5.5 - 8.0 µm of panels a1 - c1 as indicated by a red box. Likewise Figure 7, virgin ices are able to fit only some absorption due to ice, whereas models with Fluence I and II better reproduce the organics observed toward Elias 29, CRBR 2422.8-3423 and IRAS 13546-3491. On the other hand, EC 90 and EC 92 present a much more complex chemistry that cannot be fitted by the model employed in this paper. Other ices presented in Table 1 can be used to try to find a more realist ice mixture for these sources. This Figure shows again that presence of CH4 is sensitive to fluence, and therefore it can be used as a parameter to limit the chemical evolution of ices for sources like EC 90 and EC 92.

Observing the results of both models, is verified that processed ices can reproduce the absorptions profile due to ices toward YSOs better. The assignment of many bands observed toward protostars is conclusive in spite of the nature of the absorption at 6.85 µm remains ambiguous. As discussed in Schutte et al.(2003), this band is associated with the $NH_4^+$ ion. However, Zasowski et al.(2009) and Rocha et al.(2015) suggest that the absorption at 6.85 µm is associated with the $CH_3OH$ molecule. Probably, both $NH_4^+$ and $CH_3OH$ are related to the 6.85 µm absorp- tion, in different situations. For example, since the negative ion $OCN^-$ is observed in the spectrum of a protostar, this feature should be associated with $NH_4^+$, because there should be a balance of charges, as discussed by Schutte et al.(2003). On the other hand, if $OCN^-$ has not been observed, the absorption at 6.85 µm should be associated with $CH_3OH$, instead of $NH_4^+$ as pointed out by Rocha et al.(2015).Boogert et al.(2008) state that a molecule or ion associated with the feature at 6.85 µm must be produced at low temperature, by acid-base chemistry induced by UV radiation or processing by cosmic rays. Since the ices are formed in deepest regions of protoplanetary disks or dense clouds, UV photons from stellar or interstellar radiation fields cannot penetrate inside the ice-forming region (Henning et al.2013). Therefore, for many cases, the formation of the band at 6.85 µm should be associated mainly with processing of the ices by cosmic rays. Nevertheless there are evidences of exclusion of cosmic rays for evolved protostars at late class II and III (Cleeves et al.2013,2014,2015). In this way, the database presented in this paper may be used properly in astrophysical models of class 0, class I and early class II protostars. For later stages of evolution, like late class II and class III, such data should be used with caution, because the magnetic field coupled to stellar winds deflects ionizing particles and decreases the flux of cosmic rays inside protostellar environments.

## 5 CONCLUSIONS

This paper presented, for the first time, a catalog with the complex refractive index in the infrared of 13 water-containing ices bombarded by cosmic rays simulated in the laboratory at two different ion fluences. Such a database is useful to be employed into radiative transfer codes in order to reproduce in a more realistic approximation the chemical evolution of protostellar disks. As a test case, an early class II object with RADMC-3D containing grains with inclusions of 30 % and 60% of ice composed by $H_2O:CO_2:NH_3:CH_4$ was chosen. The optical depth in the infrared of the models were compared with optical depths calculated from observational YSOs.

The main conclusion from the comparisons is that processing of the ice is ubiquitous in star-forming regions. In this way data of ices energetically processed should be taken into account in radiative transfer simulations in order to reproduce the chemistry observed toward the YSOs. In fact, astrochemical models that consider the chemistry in a surface of the grains are important to understand the best pathway in production of different molecules. However, such models cannot reproduce how the chemical environment can change the absorption profile of an embedded molecule. In this case, refractive index values coming directly from the experiments are essential for astrochemical models of protostars. This conclusion is because absorption profiles due to organic compounds between 5.5 - 8.0 µm can only be reproduced in the models if processed ices are employed. However, the comparison between observational and modeled protostars shows that, there is a upper limit of flux simulated employed in laboratort that can be applied to model spectrum of protostars. Otherwise, if fluxes are very strong, formed species in moderate fluxes would be destroyed or desorbed from the ice into the gas-phase, as can be clearly seen for $CH_4$ at 7.67 µm. Moreover, the missing libration mode of water at around 12 µm observed toward CRBR 2422.8-3423 can be an evidence of abundance of $CO_2$ ice produced from processing. The chemistry observed for EC 90 nd EC 92 is much more complex that for Elias 29, CRBR 2422.8-3423 and IRAS 13546-3491, and therefore more studies should be done in order to explain the nature and abundance of organics observed toward these two objects.

Besides, the abundance of ice in the grain leads to different shapes in the absorption band of protostellar spectra, and this information can be used to infer if the grains toward YSOs are dominated by ice or by dust. Once the ice mantle formation is more evident for small silicate grains, those that present 30 % of ice in its volume are representative of objects with silicate grains larger than 0.1 µm. Indeed, the size of grains used by Rocha et al.(2015) and Pontoppidan & Dullemond(2005a), were 0.25 - 0.7 µm and 0.5 µm, respectively for Elias 29 e CRBR 2422.8-3423. On the other hand, grains containing 60 % of ice in their volume are representative for objects where the size of silicate grains is smaller than 0.1 µm. The model with 60 % of ice was representative for EC 90, EC 92 and IRAS 13546-3491.

Finally, we can conclude that chemistry driven by cosmic rays also depends on the polarity in the mixture in addition to their energy. Ices composed by polar molecules are more resistant to destruction, whereas apolar containing ices are more easily destroyed. In order to reproduce the abundance or organic compounds observed toward YSOs presented in this paper, new investigations should be done, by employing only polar and apolar ices, in addition to using another kind of ionizing radiation such as X-rays and UV ubiquitously present in star-forming regions.

**ACKNOWLEDGEMENTS**

We thank the São Paulo Research Foundation FAPESP (PhD 2013/07657-5, PD 2015/10492-3 and JP 2009/18304-0), FAPERJ (E - 110.087/2014), and the Brazilian financial agencies CNPq and CAPES. We also thank Alene Alder-Rangel for reviewing the English. The experiments were performed at GANIL and we acknowledge the important support from the CIMAP staff. In addition, we also thank the anonymous referee for a careful review of the paper.

This paper has been typeset from a TEX/LATEX file prepared by the author.